\documentclass[english]{article}

\usepackage[T1]{fontenc}
\usepackage[latin9]{inputenc}

\usepackage{cite}
\usepackage{url}
\usepackage{amsmath}
\usepackage{graphicx}

\usepackage[unicode=true]{hyperref}

\usepackage[a4paper]{geometry}
\usepackage{setspace}
\begin{document}
\title{Compartmental voter model}
\author{Aleksejus Kononovicius\thanks{email: \protect\href{mailto:aleksejus.kononovicius@tfai.vu.lt}{aleksejus.kononovicius@tfai.vu.lt};
website: \protect\url{http://kononovicius.lt}}}
\date{Institute of Theoretical Physics and Astronomy, Vilnius University}
\maketitle
\begin{abstract}
Numerous models in opinion dynamics focus on the temporal dynamics
within a single electoral unit (e.g., country). The empirical observations,
on the other hand, are often made across multiple electoral units
(e.g., polling stations) at a single point in time (e.g., elections).
Aggregates of these observations, while quite useful in many applications,
neglect the underlying heterogeneity in opinions. To address this
issue we build a simple agent--based model in which all agents have
fixed opinions, but are able to change their electoral units. We demonstrate
that this model is able to generate rank--size distributions consistent
with the empirical data.
\end{abstract}

\section{Introduction}

Most well--known models of opinion dynamics seem to imply that a
stable fixed state, either consensus or polarization, is inevitable
\cite{Galam2008ModPhysC,Castellano2009RevModPhys,Stauffer2013JStatPhys,Flache2017JASSS,Sirbu2017,Baronchelli2018RSOS,Jedrzejewski2019CRP}.
However, local and spatial heterogeneity and ongoing exchange of opinions
and cultural traits is a characterizing feature of social systems.
Various modifications of the well--known models were proposed to
account for these features, such as inflexibility \cite{Galam2007PhysA,Mobilia2007JStatMech}
or spontaneous flipping \cite{Kirman1993QJE,Granovsky1995}. Some
of the models were modified to account for the theories from the social
sciences \cite{Nail2016APPA,Duggins2017JASSS}. Effects of these modifications
are still being actively reconsidered in context of network theory,
non--linearity, complex contagion and applications towards financial
markets \cite{Kononovicius2012PhysA,Kononovicius2014EPJB,Kononovicius2014PhysA,Harmon2015Plos,Carro2016,Galam2016Chaos,Khalil2018PRE,Peralta2018Chaos,Vilela2018PhysA,Vilela2018SciRep,Artime2019,Vilela2019PhysA}.
Nevertheless even these modified models assume that opinion dynamics
occur and are observed within single electoral unit (from here on
let us also use the terms ``compartment'' and ``spatial unit'' interchangeably)
over multiple time steps. Here we propose a novel agent--based model,
which is extremely simple yet able to replicate spatial heterogeneity,
which in this paper is studied purely through socio--demographic
distributions over compartments, observed in census \cite{Schelling2006,Ausloos2007EPL,Hatna2012JASSS,Ausloos2015PhysA,Barter2019}
and electoral data \cite{Borghesi2010EPJB,Borghesi2012PLOS,FernandezGarcia2014PRL,Sano2016,Braha2017PlosOne,Fenner2017QQ,Fenner2017,Kononovicius2017Complexity,Michaud2018PRE}.

Some of the recent approaches in opinion dynamics \cite{Borghesi2010EPJB,Borghesi2012PLOS,FernandezGarcia2014PRL,Sano2016,Braha2017PlosOne,Fenner2017QQ,Fenner2017,Kononovicius2017Complexity,Michaud2018PRE,Mori2019PRE}
have combined empirical analysis of the detailed electoral data (opinions
being observed over multiple electoral units during a single time
step) with numerical modeling. Still in \cite{Sano2016,Kononovicius2017Complexity}
various groups of researchers have made the same underlying assumption:
that the electoral units are mutually independent observations from
mostly the same stationary distribution of opinions. This effectively
means that it is enough to model dynamics of a single electoral unit.
All of these approaches were based on the noisy voter model and thus
predict the opinions to be Beta distributed over time, which is somewhat
consistent with the empirical research conducted over electoral units
\cite{Hansen2003JASA,Rigdon2009APR,Kononovicius2018APPA}. In \cite{Braha2017PlosOne}
numerical modeling, using scheme similar to the one we just have described,
was supplemented by rigorous empirical analysis demonstrating strong
spatial (across US states) and temporal (over a century of US presidential
elections) correlations. These patterns could be another way of understanding
the heterogeneity of voting behavior. In \cite{Fenner2017,Fenner2017QQ}
no independency assumption was made when formulating a multiplicative
model, but the proposed model did not provide an agent--based reasoning
for the vote share heterogeneity over electoral units. Similarly \cite{Borghesi2010EPJB,Borghesi2012PLOS}
have provided a purely phenomenological fully spatial model, which
takes into account actual geospatial topology of the modeled area,
for the voter turnout rates across various countries and elections.
Only \cite{FernandezGarcia2014PRL} have built an agent--based model
with dynamics occurring across multiple electoral units. Notably the
agent--based model proposed in \cite{FernandezGarcia2014PRL} and
later improved by \cite{Michaud2018PRE} is rather complicated and
was built specifically for the elections in the United States, taking
into account topology of electoral units and empirical data of the
commuting patterns between them. Rather similar, yet much simpler,
model was proposed in \cite{Mori2019PRE}, which simply assumes that
voters can copy opinions of agents in their home and direct neighbor
nodes. This model was also able to reproduce spatial correlation patterns
considered in \cite{FernandezGarcia2014PRL}.

Here we take a similar approach to \cite{FernandezGarcia2014PRL},
but instead of commuting we consider internal migration. While commuting
may drive the short--term changes in the electoral behavior, internal
migration may give shape to the long--term trends. Furthermore in
social sciences it is quite common practice to gain insights from
the covariance between census data, which changes due to migration,
and electoral data \cite{Downes2018AJE}. Unlike \cite{FernandezGarcia2014PRL}
we do not infer the migration rates from the empirical data, the proposed
model assumes that these rates have a specific form. This allows us
to keep the model simple and focused on reproducing general patterns
of socio--demographic heterogeneity over compartments. In this sense
our approach is rather similar to the classical Schelling segregation
model \cite{Schelling2006}, but the migration rules we use are continuous
and somewhat more complicated. Similar migration rules can be found
in the classical human mobility models \cite{Barbosa2018PhysRep},
especially in the gravity model \cite{Zipf1946ASR,Jung2008EPL,Braha2011SocNet,Simini2012Nature,Pappalardo2016PCS,Barbosa2018PhysRep}
as we assume that system--wide migration rates depend on population
of source and destination compartments. Yet these models have somewhat
different goal and reproduction of the mobility patterns does not
ensure reproduction of socio--demographic heterogeneity over compartments.
Though vice versa is also true. Finally as with most models in opinion
dynamics comparison could be also drawn to the Ising model \cite{Castellano2009RevModPhys}.
Unlike the most existing approaches our model is similar not to the
Metropolis interpretation of the Ising model \cite{Metropolis1953},
but to the Kawasaki interpretation of the Ising model \cite{Kawasaki1966PR,Kawasaki1966PR2},
which assumes that particle spins are conserved, but the particles
themselves exchange places.

This paper is organized as follows. In Section~\ref{sec:model} we
introduce the compartmental model and discuss its main statistical
properties. In Section~\ref{sec:empirical} we fit the compartmental
model to a few selected empirical data sets. We finalize with the
discussion in Section~\ref{sec:conclusions}.

\section{Compartmental model\label{sec:model}}

Let us consider $N$ agents of $T$ types migrating between $M$ identical
compartments, each of which has a capacity of no more than $C$ agents.
Depending on the context the agents could be assumed to represent
residents or voters, who have certain fixed socioeconomic traits or
opinions (types). Likewise the compartments could be assumed to represent
residential areas or electoral units between which the agents can
migrate.

Let us assume that the migration rate between the compartments $i$
and $j$ for the agents of type $k$ has the following form:
\begin{equation}
\lambda_{i\rightarrow j}^{\left(k\right)}=\begin{cases}
X_{i}^{\left(k\right)}\left(\varepsilon^{\left(k\right)}+X_{j}^{\left(k\right)}\right) & \text{if }i\neq j\text{ and }N_{j}<C,\\
0 & \text{otherwise}.
\end{cases}\label{eq:model}
\end{equation}
In the above $X_{i}^{\left(k\right)}$ is the number of agents of
type $k$ in the compartment $i$, $N_{j}$ is the number of agents
of all types in the compartment $j$. The migration rate is composed
of two terms. One of the terms linearly depends on the number of agents
present in the source compartment $X_{i}^{\left(k\right)}$. This
term represents spontaneous (idiosyncratic) migration and $\varepsilon^{\left(k\right)}$
is the relative spontaneous (idiosyncratic) migration rate for the
agents of type $k$. The second term involves the number of agents
present in the source compartment $X_{i}^{\left(k\right)}$ and the
number of agents present in the target compartment $X_{j}^{\left(k\right)}$.
This nonlinear term represents migration induced by the interaction
processes such as recruitment or homophily. To keep the migration
rates as simple as possible we assume that such interactions are possible
only between the agents of the same type. Induced migration is assumed
to occur at a unit rate. Note that transition rates of the same form
are present in the noisy voter model. This is why we consider the
compartmental model to belong to the voter model family, ignoring
the fact the compartmental model does not describe any actual opinion
dynamics.

Here we would like to draw clear distinction between two different
distributions, which could be obtained from the series of $X_{i}^{\left(k\right)}\left(t\right)$
and other related variables. Usually in theoretical sociophysics papers
spatial index $i$ is fixed, while the observations are made over
multiple different time steps $t$. If the model is ergodic (this
model is ergodic as well as many other well--known models) and the
series is long enough, then the sample distribution should approach
the stationary distribution. Thus we will refer to this sample distribution
as the stationary distribution and use the probability mass function
(abbr. PMF) or the probability density function (abbr. PDF), $p\left(X_{i}^{\left(k\right)}\right)$,
to show the results. Unlike most of the previous approaches this model
allows for $i$ to be variable and $t$ to be fixed. In this case
the sampling is no longer temporal, but instead is made over compartments.
Therefore we refer to this sample distribution as the compartmental
distribution and use the compartmental rank--size distribution (abbr.
CRSD), $\tilde{X}_{r}^{\left(k\right)}$, to show the results. If
compartments would be mutually independent, then the compartmental
distribution should also approach the stationary distribution. Yet
the compartments in this model are not independent as the total number
of agents is fixed and agent leaving one compartment must move to
some other compartment.

Although the compartmental model is quite simple, it is not straightforward
to obtain closed form expressions for the stationary distribution
or the compartmental distribution. If the compartments are able to
hold all the agents, $C=N$, it is quite straightforward to derive
the total entry and exit rates for agents of type $k$:
\begin{align}
\lambda_{i+}^{\left(k\right)} & =\sum_{j=1}^{M}\lambda_{j\rightarrow i}^{\left(k\right)}=\left[N^{\left(k\right)}-X_{i}^{\left(k\right)}\right]\left(\varepsilon^{\left(k\right)}+X_{i}^{\left(k\right)}\right),\label{eq:model-entry}\\
\lambda_{i-}^{\left(k\right)} & =\sum_{j=1}^{M}\lambda_{i\rightarrow j}^{\left(k\right)}=X_{i}^{\left(k\right)}\left(\left[M-1\right]\varepsilon^{\left(k\right)}+\left[N^{\left(k\right)}-X_{i}^{\left(k\right)}\right]\right).\label{eq:model-exit}
\end{align}
In the above $N^{\left(k\right)}=\sum_{i=1}^{M}X_{i}^{\left(k\right)}$
is the total number of agents of type $k$. These rates are identical
to the transition rates of the multi--state noisy voter model \cite{Kononovicius2017Complexity}.
This similarity allows us to conclude that $x_{i}^{\left(k\right)}=X_{i}^{\left(k\right)}/N^{\left(k\right)}$
will be Beta distributed in the asymptotic limit, while as whole the
value sets $\vec{x}^{\left(k\right)}$ will be Dirichlet distributed.
If $N^{\left(k\right)}$ is finite, then $X_{i}^{\left(k\right)}$
will follow Beta-binomial distribution. We can confirm this intuition
from the detailed balance condition (which holds for $X_{i}^{\left(k\right)}\in\left[0,N^{\left(k\right)}-1\right]$):
\begin{equation}
p\left(X_{i}^{\left(k\right)}\right)\lambda_{i+}^{\left(k\right)}\left(X_{i}^{\left(k\right)}\right)=p\left(X_{i}^{\left(k\right)}+1\right)\lambda_{i-}^{\left(k\right)}\left(X_{i}^{\left(k\right)}+1\right).\label{eq:detailed-balance}
\end{equation}
Rearranging detailed balance condition gives us a set of recursive
equations:
\begin{equation}
p\left(X_{i}^{\left(k\right)}+1\right)=\frac{\left[N^{\left(k\right)}-X_{i}^{\left(k\right)}\right]\left(\alpha+X_{i}^{\left(k\right)}\right)}{\left(X_{i}^{\left(k\right)}+1\right)\left(\beta+\left[N^{\left(k\right)}-X_{i}^{\left(k\right)}+1\right]\right)}p\left(X_{i}^{\left(k\right)}\right).
\end{equation}
In the above we have used $\alpha$ in place of $\varepsilon^{\left(k\right)}$
and $\beta$ in place of $\left[M-1\right]\varepsilon^{\left(k\right)}$.
We can see that PMF of the Beta-binomial distribution:
\begin{equation}
p\left(X_{i}^{\left(k\right)}\right)=C\frac{B\left(X_{i}^{\left(k\right)}+\alpha,N^{\left(k\right)}-X_{i}^{\left(k\right)}+\beta\right)}{B\left(\alpha,\beta\right)}
\end{equation}
satisfies the recursive equations. In the above $C$ is normalization
constant and $B\left(x,y\right)$ is the Beta function. Identical
derivation of the stationary PMF for the noisy voter model can be
found in \cite{Mori2019PRE}.

\begin{figure}
\begin{centering}
\includegraphics[width=0.7\textwidth]{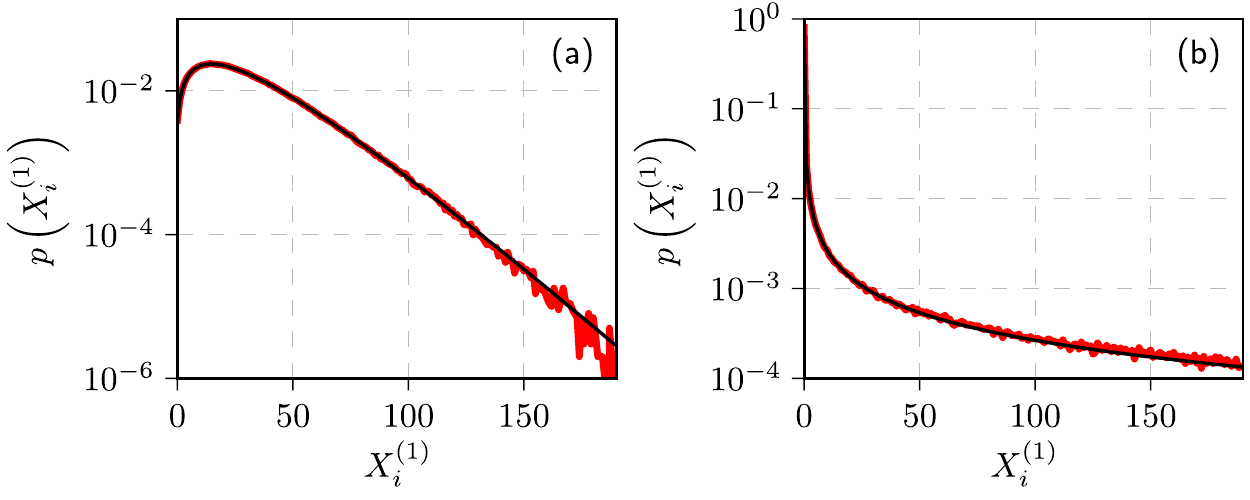}
\par\end{centering}
\caption{(color online) Comparison between the numerical PMFs of $X_{i}^{\left(1\right)}$
(red curves) against the Beta-binomial PMFs (black curves). Model
parameter values: $N=3000$, $T=1$, $M=100$ and $C=N$ (in all cases),
$\varepsilon^{\left(1\right)}=2$ (a) and $0.03$ (b). Assumed parameter
values of the Beta-binomial distribution: $N=3000$, $\alpha=\varepsilon^{\left(1\right)}$
and $\beta=\left(M-1\right)\varepsilon^{\left(1\right)}$ (in all
cases).\label{fig:xpmf}}
\end{figure}

\begin{figure}
\begin{centering}
\includegraphics[width=0.7\textwidth]{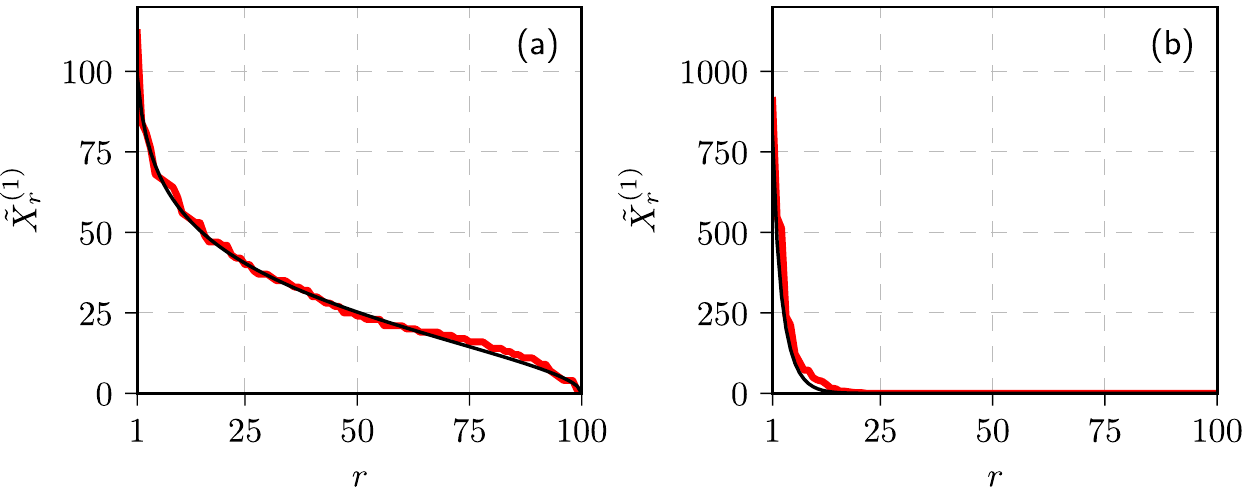}
\par\end{centering}
\caption{(color online) Comparison between the numerical CRSDs of $X_{i}^{\left(1\right)}$
(red curves) against the Beta-binomial RSDs (black curves). Numerical
CRSDs were obtained from the same simulations as in Fig.~\ref{fig:xpmf}.
Assumed parameter values of the Beta-binomial distribution are also
the same as in Fig.~\ref{fig:xpmf}.\label{fig:xrsd}}
\end{figure}

In Fig.~\ref{fig:xpmf} we can see that Beta-binomial PMF fits the
numerical PMF rather well. While Beta-binomial RSD seems to be a good
fit for numerical CRSD as shown in Fig.~\ref{fig:xrsd}.

\begin{figure}
\begin{centering}
\includegraphics[width=0.7\textwidth]{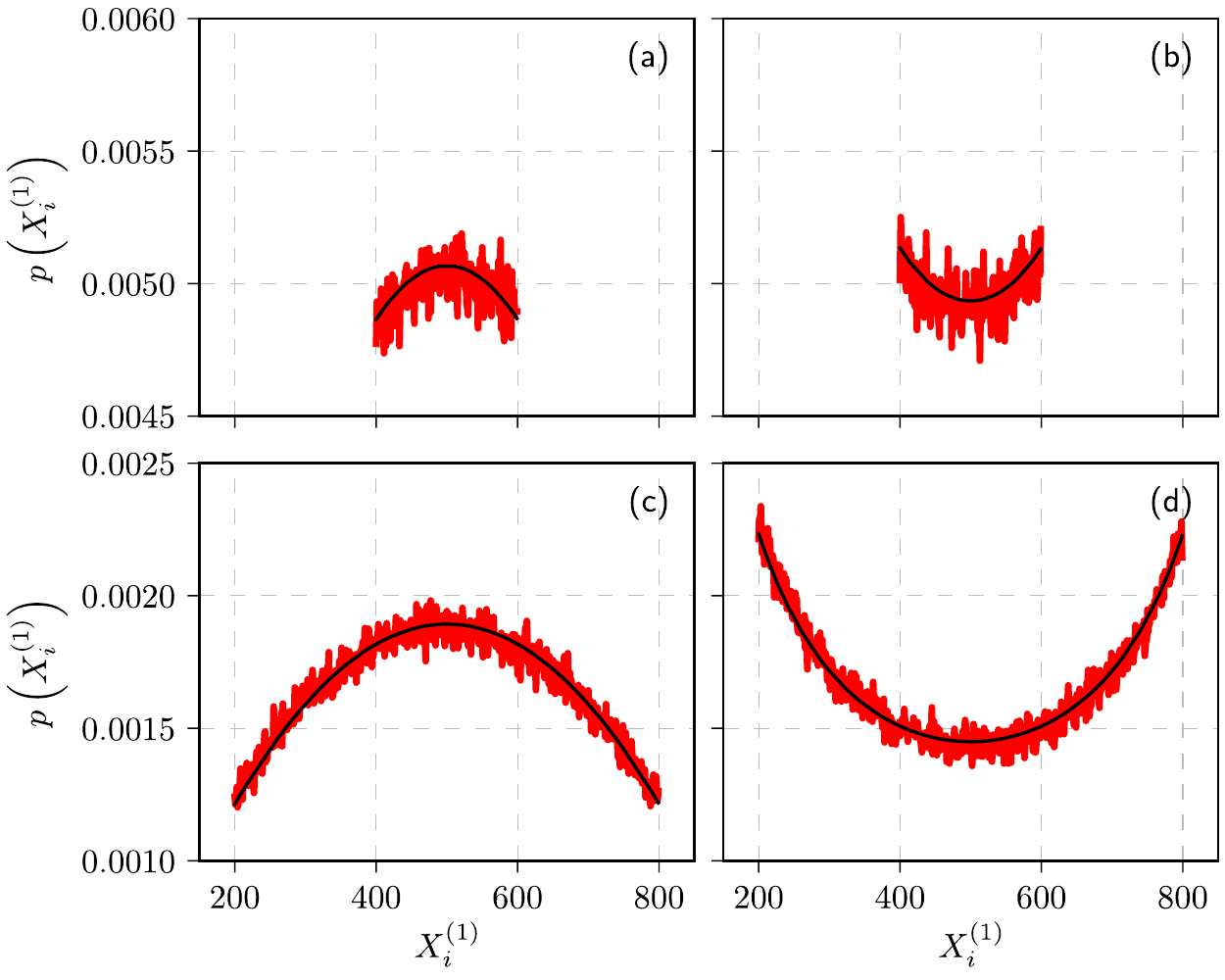}
\par\end{centering}
\caption{(color online) Comparison between the numerical PMFs of $X_{i}^{\left(1\right)}$
(red curves) against the truncated Beta-binomial PMFs (black curves).
Model parameter values: $N=1000$, $T=1$ and $M=2$ (in all cases),
$C=600$ ((a) and (b)) and $800$ ((c) and (d)), $\varepsilon^{\left(1\right)}=2$
((a) and (c)) and $0.03$ ((b) and (d)). Assumed parameter values
of the truncated Beta-binomial distribution: $N=1000$, $\alpha=\varepsilon$
and $\beta=\left(M-1\right)\varepsilon$, while allowing for $X_{i}^{\left(1\right)}\in\left[N-C,C\right]$
(in all cases).\label{fig:xpmf-cap-t1m2}}
\end{figure}

If $\frac{N}{M}<C<N$, we can no longer ignore the fact the number
of agents within compartment, $N_{i}$, is capped. Consequently we
can no longer simplify the sums in the total entry and exit rates,
Eqs.~(\ref{eq:model-entry}) and (\ref{eq:model-exit}), to a tractable
form even for the simplest cases. Though it is possible to obtain
stationary PMF for the specific cases by treating the model as one
dimensional Markov chain or from detailed balance condition. Using
the Markov chain approach we were able to find that for $T=1$ and
$M=2$ PMF of $X_{i}^{\left(k\right)}$ precisely follows truncated
Beta-binomial distribution (see Fig.~\ref{fig:xpmf-cap-t1m2}). One
can easily confirm this from detailed balance condition Eq.~(\ref{eq:detailed-balance}),
which would now apply to a narrower interval $X_{i}^{\left(k\right)}\in\left[C,N^{\left(k\right)}-C-1\right]$.
As the condition itself has not changed, in order to get stationary
PMF in this case we just have to appropriately truncate the Beta-binomial
PMF. Yet for other, more complicated cases, we were unable to discover
a general pattern with either of the discussed approaches. Furthermore
size of the Markov chain (the number of states) seems to grow exponentially
with both $T$ and $M$. The number of recursive equations to be satisfied
also seems to grow exponentially fast. Hence for realistic $N$, $T$
and $M$ it is not feasible to obtain an analytical result.

\begin{figure}
\begin{centering}
\includegraphics[width=0.7\textwidth]{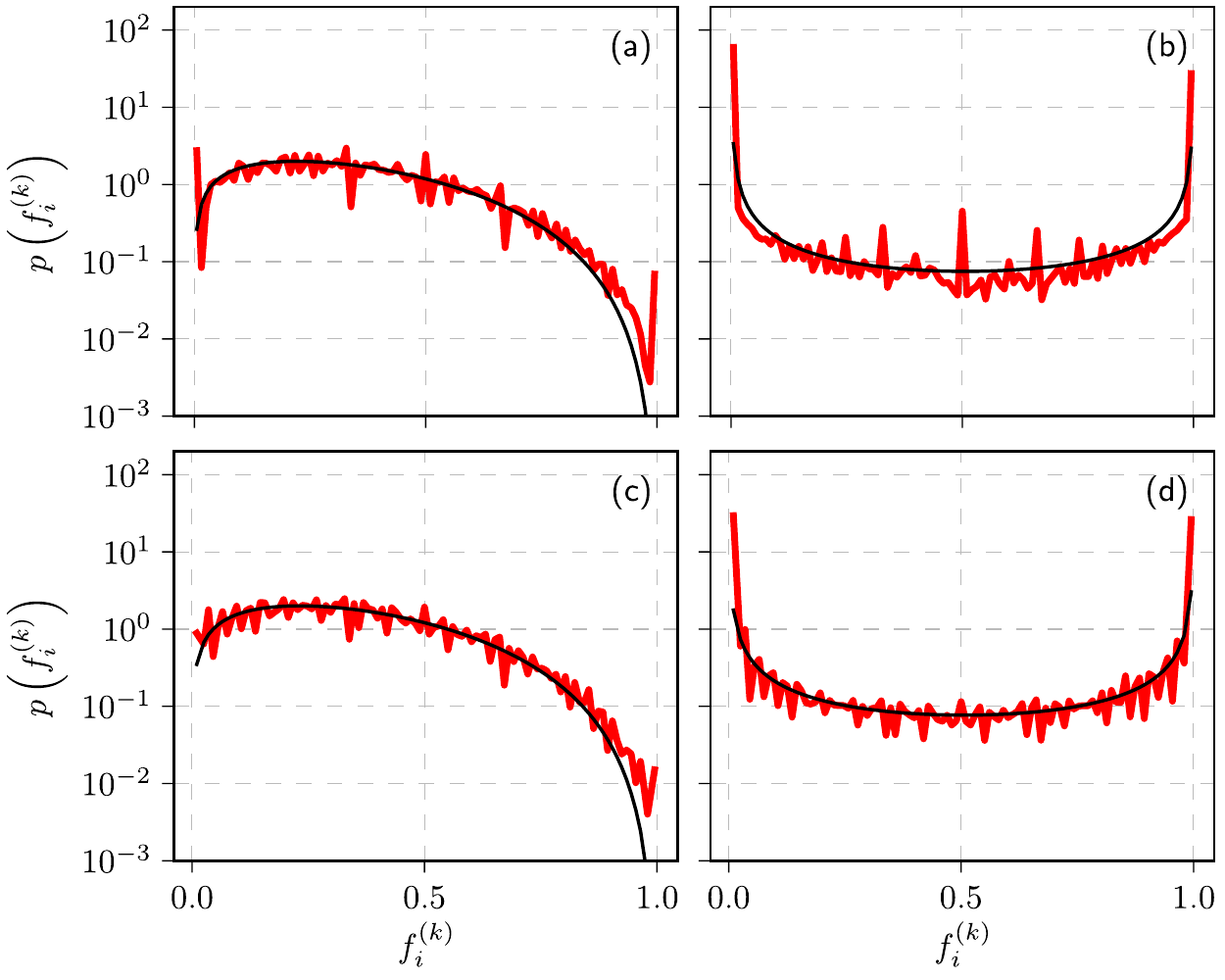}
\par\end{centering}
\caption{(color online) Fitting the numerical PDFs of $f_{i}^{\left(k\right)}$
(red curves) using PDF of Beta distribution (black curves). Model
parameter values: $N^{\left(1\right)}=N^{\left(2\right)}=N^{\left(3\right)}=1000$
(there were $N=3000$ agents in total), $M=100$, $T=3$ and $\varepsilon^{\left(1\right)}=\varepsilon^{\left(2\right)}=\varepsilon^{\left(3\right)}=\varepsilon$
(in all cases), $C=N$ ((a) and (b)) and $C=35$ ((c) and (d)), $\varepsilon=2$
((a) and (c)) and $0.03$ ((b) and (d)). Best fit parameter values
for the Beta distribution: (a) $\alpha=1.69$ and $\beta=3.46$, (b)
$\alpha=0.031$ and $\beta=0.055$, (c) $\alpha=1.76$ and $\beta=3.53$,
(d) $\alpha=0.033$ and $\beta=0.055$.\label{fig:frac-fig}}
\end{figure}

In this model we have assumed that compartments are identical, while
in the real world the compartments will not be identical. In the real
world there will be a natural variation in number of people in spatial
units, due to historical or geographical reasons. Also number of people
in empirical data sets might be larger than numerical simulations
could deal with (at least in reasonable computation time). For these
reasons instead of making direct comparisons with the raw $X_{i}^{\left(k\right)}$,
we introduce use a scaled variable, which we refer to as a population
fraction:
\begin{equation}
f_{i}^{\left(k\right)}=\frac{X_{i}^{\left(k\right)}}{N_{i}}.
\end{equation}
In the empirical context closest match to the population fraction
would be the vote share, which in empirical works is defined as fraction
of votes case for specific party or candidate during the given election.
Obtaining closed form expression for the stationary distribution of
$f_{i}^{\left(k\right)}$ is quite problematic, because even in the
simplest case $f_{i}^{\left(k\right)}$ is a ratio of a Beta distributed
random variable and a sum of correlated Beta distributed random variables.
In mathematical statistics some results are known only for a very
simple combinations of independent Beta random variables \cite{Pham1994,Pham2000}.
Nevertheless at least for a few selected cases we see that $f_{i}^{\left(k\right)}$
has a stationary distribution which is well approximated by the Beta
distribution (see Fig.~\ref{fig:frac-fig}), although the distribution
parameter values must be fitted on case--by--case basis and some
discrepancies are noticeable. Though notably the fitted parameter
values for $f_{i}^{\left(k\right)}$, we have used Maximum Likelihood
Estimation to obtain them, appear to be reasonably close to $\alpha=\varepsilon^{\left(k\right)}$
and $\beta=\left(\sum_{t=1}^{T}\varepsilon^{\left(t\right)}\right)-\varepsilon^{\left(k\right)}$.

\section{Comparison with selected empirical data sets\label{sec:empirical}}

In this section we use the compartmental model to fit the empirical
census and electoral data. We take three subsets of UK census 2011
data (which can be obtained from the NOMIS website\footnote{\url{https://www.nomisweb.co.uk/query/select/getdatasetbytheme.asp}})
and one subset of Lithuanian parliamentary election 1992 data (which
can be obtained from a dedicated GitHub repository\footnote{\url{https://github.com/akononovicius/lithuanian-parliamentary-election-data}}).
These subsets were selected semi--randomly, namely we have ensured
that all considered ``groups'' of people would be quite well represented
and reasonably segregated. We have rejected some other randomly selected
subsets, because they were either dominated by a single ``group''
or were too uniformly spread out.

\begin{figure}
\begin{centering}
\includegraphics[width=0.7\textwidth]{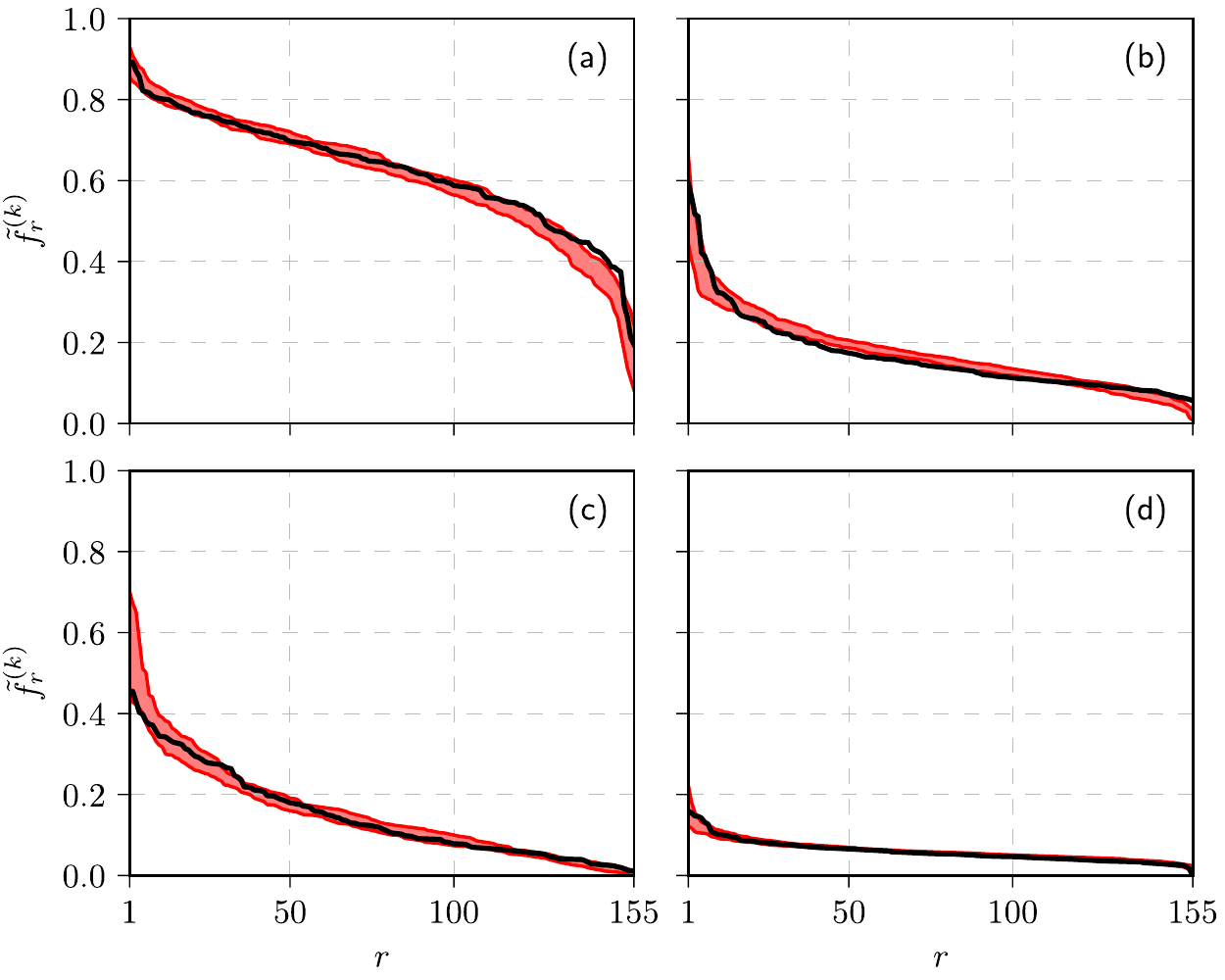}
\par\end{centering}
\caption{(color online) Comparison between the numerical CRSDs of $f_{i}^{\left(k\right)}$
(red shaded areas) against empirical CRSDs for the ethnic group distributions
in London (black curves). Different sub--figures show the curves
for the different considered ethnic groups: (a) White (index $w$),
(b) Asian (index $a$), (c) Black (index $b)$ and (d) other (index
$o$). Numerical results are reported using $95\%$ confidence interval,
which spans the red shaded area. Model parameter values: $N^{\left(w\right)}=48515$,
$N^{\left(a\right)}=12865$, $N^{\left(b\right)}=11470$ and $N^{\left(o\right)}=4495$
(there were $N=77345$ agents in total), $\varepsilon^{\left(w\right)}=2.5$,
$\varepsilon^{\left(a\right)}=4$, $\varepsilon^{\left(b\right)}=1.5$,
$\varepsilon^{\left(o\right)}=15$, $M=155$, $C=600$.\label{fig:london-race}}
\end{figure}

The first example we consider is the ethnic group distribution over
the postal districts (in total $155$ of them) in London (see Fig.~\ref{fig:london-race}).
We consider three most well represented ethnic groups (White, Asian
and Black), while combining less represented groups in to the other
group. Members of mixed ethnic groups were assigned to either Asian,
Black or the other group. In the simulation we have used $77345$
agents to represent $4856091$ people in the data set (approximately
$1:60$ ratio). Other model parameter values are listed in the caption
of Fig.~\ref{fig:london-race}.

\begin{figure}
\begin{centering}
\includegraphics[width=0.95\textwidth]{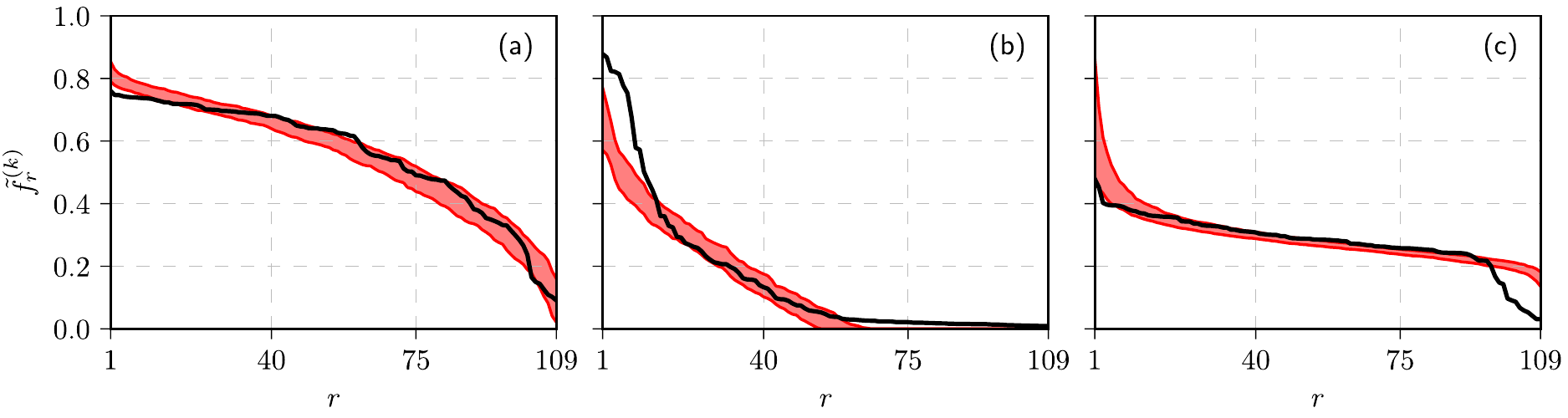}
\par\end{centering}
\caption{(color online) Comparison between the numerical CRSDs of $f_{i}^{\left(k\right)}$
(red shaded areas) against empirical CRSDs for the religious group
distributions in Leicester (black curves). Different sub--figures
show the curves for the different considered religious groups: (a)
Christians (index $c$), (b) no religion (index $n$) and (c) other
(index $o$). Numerical results are reported using $95\%$ confidence
interval, which spans the red shaded area. Model parameter values:
$N^{\left(c\right)}=30411$, $N^{\left(n\right)}=8829$ and $N^{\left(o\right)}=15151$
(there were $N=54391$ agents in total), $\varepsilon^{\left(c\right)}=2.5$,
$\varepsilon^{\left(n\right)}=0.01$, $\varepsilon^{\left(o\right)}=50$,
$M=109$, $C=600$.\label{fig:leicester-religion}}
\end{figure}

The second example we consider is the religious group distribution
over the postal sectors ($109$ of them) in Leicester (see Fig.~\ref{fig:leicester-religion}).
We consider three groups: Christian, no religion and other religion.
These groups were selected, because they were the most well represented.
No religion other than Christianity was sufficiently well represented
so all their followers were combined into the other religion group.
In the simulation we have used $54391$ agents to represent $925071$
people in the data set (approximately $1:17$ ratio). Other model
parameter values are listed in the caption of Fig.~\ref{fig:leicester-religion}.

\begin{figure}
\begin{centering}
\includegraphics[width=0.7\textwidth]{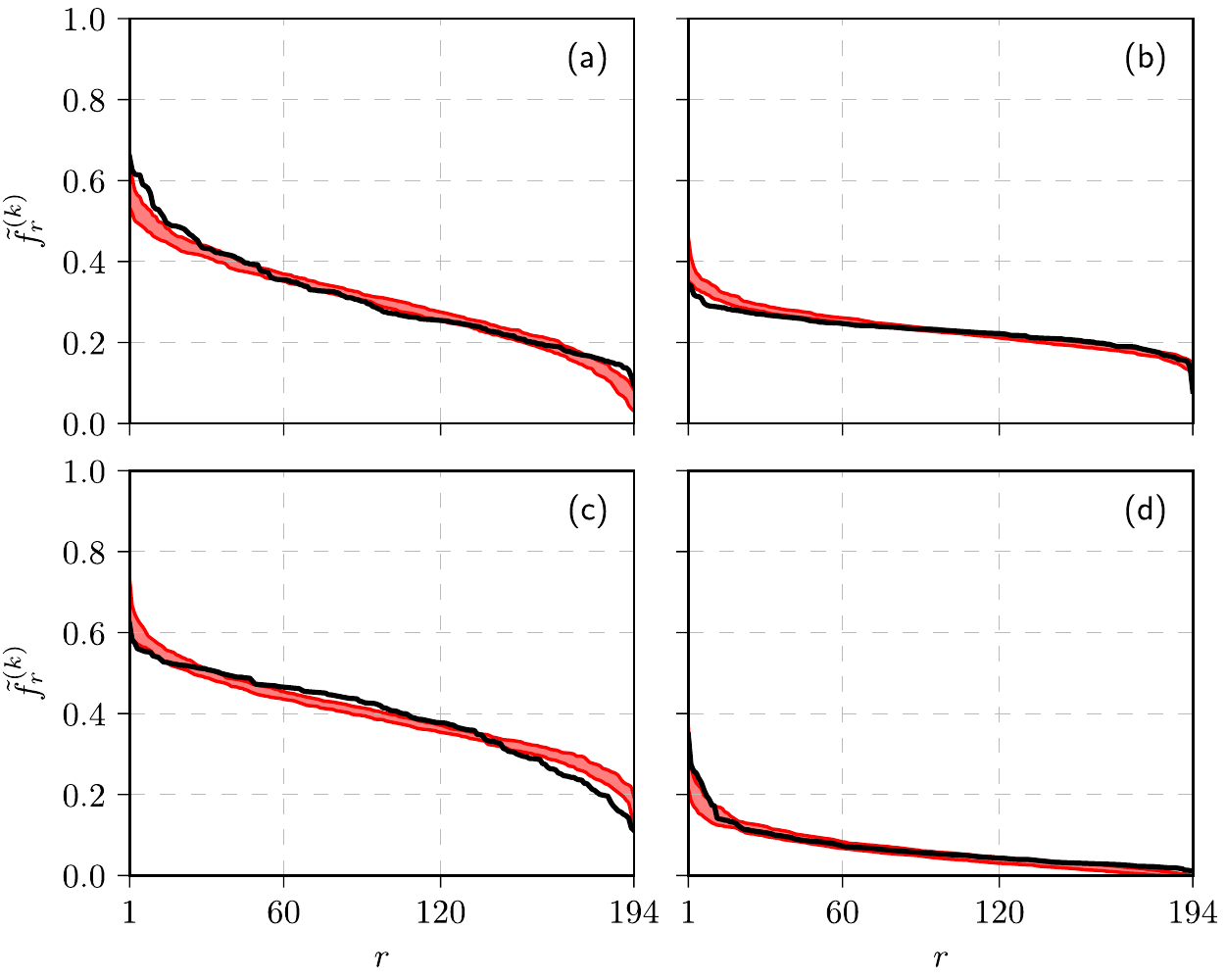}
\par\end{centering}
\caption{(color online) Comparison between the numerical CRSDs of $f_{i}^{\left(k\right)}$
(red shaded areas) against empirical CRSDs for the NS--SEC class
distributions in Sheffield (black curves). Different sub--figures
show the curves for the different considered NS--SEC classes: (a)
higher occupations (index $1$), (b) intermediate occupations (index
$2$), (c) lower occupations (index $3$) and (d) unemployed (index
$u$). Numerical results are reported using $95\%$ confidence interval,
which spans the red shaded area. Model parameter values: $N^{\left(1\right)}=29876$,
$N^{\left(2\right)}=22310$, $N^{\left(3\right)}=38218$ and $N^{\left(u\right)}=6596$
(there were $N=97000$ agents in total), $\varepsilon^{\left(1\right)}=3$,
$\varepsilon^{\left(2\right)}=50$, $\varepsilon^{\left(3\right)}=12$,
$\varepsilon^{\left(u\right)}=2$, $M=194$, $C=600$.\label{fig:sheffield-employment}}
\end{figure}

The third and final UK census example is based on the National Statistics
Socio--Economic Classification (abbr. NS--SEC). UK census 2011 data
is available in the eight--class resolution, but we down scale to
the three--class resolution to ensure that each class would be well
represented. This gives us four groups: higher (class 1), intermediate
(class 2), lower occupations (class 3) and long--term unemployed.
We have ignored data regarding persons still in education and ones
who have retired. In the simulation we have used $97000$ agents to
represent $902970$ people in the data set (approximately $1:9$ ratio).
In Fig.~\ref{fig:sheffield-employment} we have plotted the employment
class distribution over the postal sectors ($194$ of them) in Sheffield.
Caption of the figure lists other parameter values used in the simulation.

\begin{figure}
\begin{centering}
\includegraphics[width=0.7\textwidth]{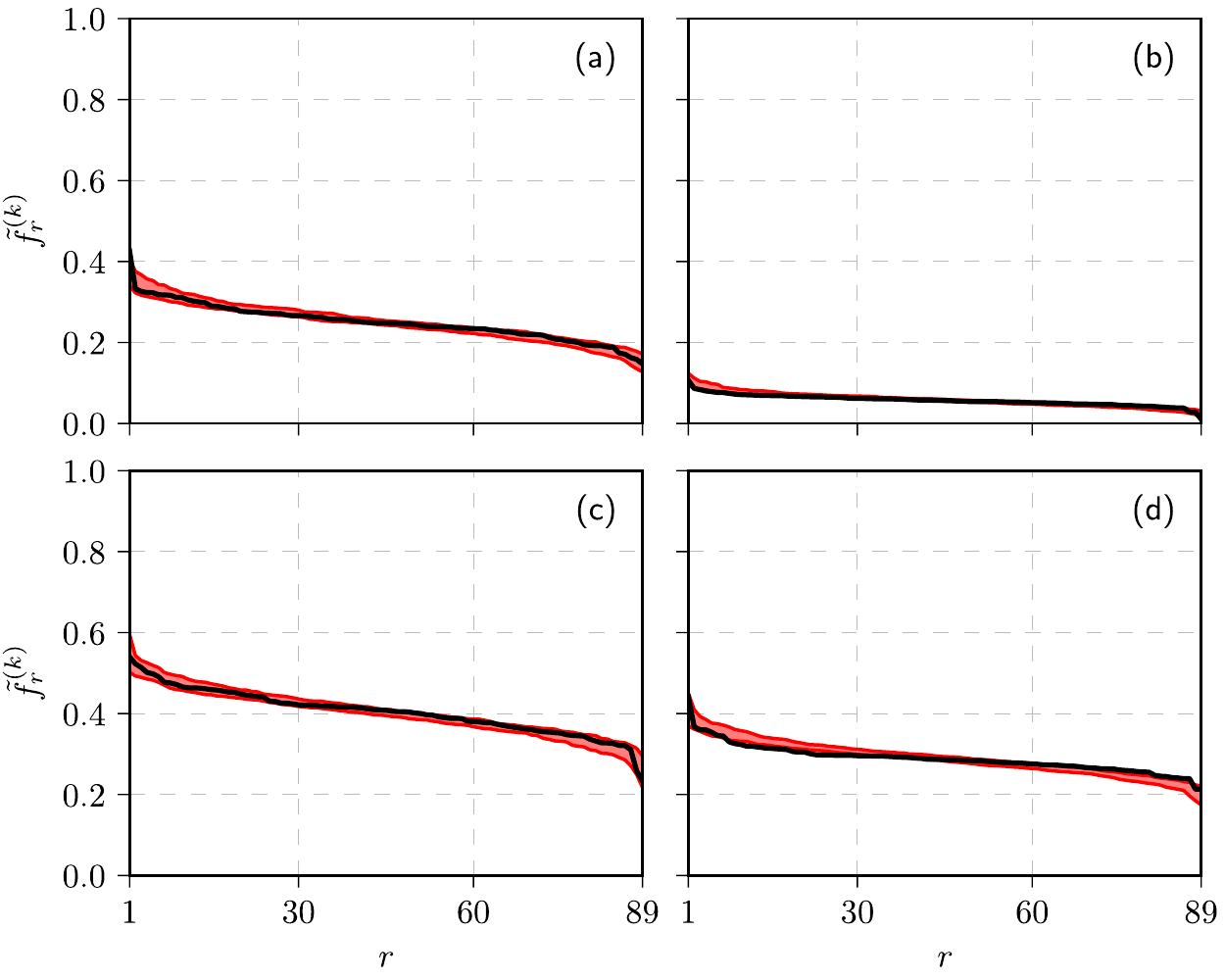}
\par\end{centering}
\caption{(color online) Comparison between the numerical CRSDs of $f_{i}^{\left(k\right)}$
(red shaded areas) against empirical vote share CRSDs observed in
Vilnius during Lithuanian parliamentary election of 1992 (black curves).
Different sub--figures show the curves for the different considered
parties: (a) S\k{a}j\={u}džio koalicija (index $s$), (b) Lietuvos
krikš\v{c}ioni\k{u} demokart\k{u} partija (index $l$), (c) Lietuvos
demokratin\.{e} darbo partija (index $d$) and (d) other parties (index
$o$). Numerical results are reported using $95\%$ confidence interval,
which spans the red shaded area. Model parameter values: $N^{\left(s\right)}=11125$,
$N^{\left(l\right)}=2581$, $N^{\left(d\right)}=17978$ and $N^{\left(o\right)}=12816$
(there were $N=44500$ agents in total), $\varepsilon^{\left(s\right)}=\varepsilon^{\left(l\right)}=\varepsilon^{\left(d\right)}=25$,
$\varepsilon^{\left(o\right)}=75$, $M=89$, $C=600$.\label{fig:lt-elections}}
\end{figure}

For the last example we take a subset of Lithuanian parliamentary
election 1992 data: the vote share distribution over the polling stations
($89$ of them) in Vilnius (see Fig.~\ref{fig:lt-elections}). As
was done earlier in \cite{Kononovicius2017Complexity} we have selected
three most successful parties to have their own groups, while all
other less successful parties were combined into a single group. In
the simulation we have used $44500$ agents to represent $177505$
voters in the data set (approximately $1:4$ ratio). Other model parameter
values are listed in the caption of Fig.~\ref{fig:lt-elections}.

As can be seen in the figures the compartmental model despite its
simplicity is able to provide a rather good fits for the empirical
data. Some deviations are observed due to variety of factors. Namely,
we have neglected in--type heterogeneity, while not all people of
the same ``group'' would have the same tendency to migrate independently.
Also not all ``classifications'' of the people are equally important
for the migration purposes (e.g., sex would be not important at all)
or multiple ``classifications'' might play a role at the same time
(e.g., if the person strongly identifies with both his ethnicity and
religious beliefs).

\section{Conclusions\label{sec:conclusions}}

In the \cite{FernandezGarcia2014PRL} Fernandez-Garcia and coauthors
have asked whether the voter model is a model for voters. Here we
have proposed a simple compartmental model based on the noisy voter
model with the aim to provide our answer to this question. We have
assumed that agents (voters) have fixed types (cultural traits or
opinions), but are able to migrate between the compartments (residential
areas or electoral districts). In such model there is no actual opinion
dynamics only spatial organization (migration). While exploring the
general statistical properties of the proposed model we have shown
that in some cases the model generates Beta distributed random variables,
which is consistent with the empirical observations. To strengthen
our argument we have used the compartmental model to provide a rather
good fits for the empirical census and electoral data. We believe
that this allows us to conclude that the spatial variations in the
electoral data can arise purely from the spatial organization patterns.
So, while the voter model is assumed to be a model for voters similar
patterns can be recovered even without any actual opinion dynamics.
Alternatively, the actual opinion exchange process could be described
by a different model, maybe even a convergent model.

The proposed model is quite simple and invites variety of further
explorations both from numerical and analytical perspectives. From
numerical perspective it would reasonable to give the compartments
some actual spatial structure and explore the arising spatial correlations.
This would also allow to make comparisons to the human mobility models.
While from analytical perspective it would be quite useful to establish
a technique to derive the compartmental rank--size distributions,
which emerge after an almost infinite time. From both perspectives
an important future consideration would be to allow the agents to
change their ``types'' and see what dynamics arise from the two competing
processes (compartmental organization and opinion exchange).

\section*{Acknowledgements}

Research was funded by European Social Fund (Project No 09.3.3-LMT-K-712-02-0026).

\end{document}